\documentclass{ws-procs10x7}
\usepackage{balance}
\usepackage{psfig}
 
\newcolumntype{d}[1]{D{.}{.}{#1}}

\def\a{\alpha}
\def\N{{\cal N}}

\makeindex
\begin{document}

\title{STRINGS FOR QUANTUMCHROMODYNAMICS}

\author{VOLKER SCHOMERUS}

\address{DESY Theory group, DESY Hamburg, D - 22603 Hamburg, Germany\\
 $^*$E-mail: volker.schomerus@desy.de}

\twocolumn[\maketitle\abstract{During the last decade, intriguing
dualities between gauge and string theory have been found and
explored. They provide a novel window on strongly coupled gauge
physics, including QCD-like models. Based on a short historical
review of modern string theory, we shall explain how so-called
AdS/CFT dualities emerged at the end of the 1990s. Some of their
concrete implications and remarkable recent progress are then
illustrated for the simplest example, namely the multicolor limit
of $\N=4$ SYM theory in four dimensions. We end with a few comments
on existing extensions to more realistic models and applications, in
particular to the sQGP. This text is meant as a non-technical
introduction to gauge/string dualities for (particle) physicists.}
\vspace*{-4mm} 
]

\section{Introduction and early history}

String theory today is mostly perceived as a theory of Planck
scale physics, offering one promising path towards a unification
of all interactions. But when it was first born around 1970, it
was meant to model strong interactions, in particular the large
number of resonances that were observed in laboratory
experiments. It is widely known that these early applications of
fundamental strings to GeV scale physics failed rather miserably.
Therefore, it may seem a bit surprising to observe the large
number of recent meetings devoted to connections of strings and
Quantumchromodynamics (QCD) or even to see string theory being
highlighted in talks on QCD and Heavy Ion collisions at this
conference (see in particular the contributions of
G.\ Marchesini\cite{Marchesini:2006rg} and X.N.\ Wang).

The aim of this lecture is to explain how such a turn could occur
and to give some idea of what we can expect from it in the future.
To this end, we shall retrace the history of string theory,
starting from the early attempts to describe strong interactions,
then passing through more modern developments in the 1980s and
1990s until we reach the discovery of intriguing novel dualities
between string and gauge theory that became known
as AdS/CFT correspondence.

During the 1960s, physicists found an enormous number of strongly
interacting hadrons. The longer the searches were pursued, the
higher became the spins $J$ and masses $m$ of the observed
resonances. In addition, a curious linear relation $J = \a_0 +
\a' m^2$ emerged which could be characterized by the so-called Regge
slope $\a'$ and intercept $\a_0$. In the absence of any other
theoretical explanation, string theory seemed to provide an
exciting perspective on these findings. Namely, it was shown
that simple (open) string theories in flat
space\cite{Nambu:1970,Nielsen:1970,Susskind:1970xm} naturally
lead to a scattering amplitudes that had been proposed by
Veneziano\cite{Veneziano:1968yb}
\begin{equation} \label{Veneziano}
{\cal A}(s,t) \ = \ \frac{\Gamma(-1-\a's)\Gamma(-1-\a't)}
{\Gamma(-2 -\a' (s+t))}
\end{equation}
where $s,t$ are the Mandelstam invariants of the scattering
process. While $s$ parametrizes the center an mass energy, $t$ is
related to the scattering angle of the event. From the pole
structure of $\Gamma$ functions it is easy to deduce the following
expansion of ${\cal A}$ at small $s$,
$$
{\cal A}(s,t) \ \sim \ - \sum_J \ \frac{P^J(s)}{\a' t - J + 1}\ . \
\ \ \ \ \mbox{(small s)}
$$ Here, $P^J$ is a polynomial of degree $J$. Hence, ${\cal A}$
does indeed encode the exchange of resonances which lie on a
Regge trajectory $m^2 = (J-1)/\a'$. This success of string theory
is not too difficult to understand. String modes in flat space are
harmonic oscillators and it is well known from basic quantum
mechanics that these possess a linear spectrum with a distance
between the spectral lines that is determined by the tension $T_s$
of the string. If we choose the latter to be $T_s \sim 1/\a'$ then
we may identify hadronic resonances with vibrational modes of a
string (provided we are willing to close an eye on the first
resonance with $J=0$ which is tachyonic).

Obviously, the formula (\ref{Veneziano}) must not be restricted to
small center of mass energies. It can also be evaluated e.g.\ for
fixed angle scattering at large $s$. Using once more some simple
properties of the $\Gamma$ function one can derive
$$ {\cal A}(s,t) \ \sim\ f(\theta)^{-1-\a's}\ \ \ \
\mbox{(large s)}
$$
where $f$ is some function of the center of mass scattering angle
$\theta$ whose precise form is not relevant for us. The result
shows that fixed angle scattering amplitudes predicted by flat
space string theory fall of exponentially with the energy $\sqrt
s$. Unfortunately for early string theory, this is not at all what
is found in experiments which display much harder high energy
cross sections. The failure of string theory to produce the
correct high energy features of scattering experiments is once
more easy to understand: strings are extended objects and as such
they do not interact in a single point but rather in an extended
region of space-time. Consequently, their scattering amplitudes
are rather soft at high energies (small distances), at least
compared to point particles. In this sense, experiments clearly
favored a point particle description of strongly interacting
physics over fundamental GeV scale strings.

As we all know, a highly successful point particle model for
strong interactions, known as Quantumchromodynamics, was
established only a few years later. It belongs to the class
of gauge theories that have ruled
our description of nature for several decades now. Due to its
asymptotic freedom, high energy QCD is amenable to perturbative
treatment. On the other hand, low energy (large distance) physics
is strongly coupled and therefore remains difficult to address.
Even though the problem to understand e.g.\ confinement remains
unsolved, QCD has at least never made any predictions that could
be clearly falsified in a simple laboratory experiment, in
contrast to what we have reviewed about early string theory. So,
in spite of its intriguing success with hadronic resonances, string
theory retracted from the area of strong interactions, it even
disappeared from physics for more than a decade before re-emerging
as a quantum theory of gravity.

Our discussion throughout the last few paragraphs did provide a
very simple explanation for the failure of early string theory,
showing that it was linked directly to the strings' extended
nature. This might make it difficult to believe that string theory
could ever make it back into strong interactions. But the early
attempts were based on the implicit assumption that the relevant
strings were moving in the same 4-dimensional space-time as the
gauge theory objects (possibly with some additional compact
dimensions). At the time, there was neither any reason nor
sufficient technical ability to think of any other scenario.
But as string theory was developed, the assumption appeared
less and less natural until it was eventually understood that
many gauge theories do admit a dual description that involves
strings moving in curved 5-dimensional backgrounds.%

\section{A Sketch of String Theory}
In order to prepare for such insights, we need to review the
development of string theory throughout the 1980s and 90s. We
shall begin with a brief sketch of the relation between closed
strings and gravity, then discuss the so-called branes along with
their open string excitations. The latter bring in gauge theories
and thereby shall enable us to argue for an intriguing novel
relation between closed strings and gauge theory.

\subsection{Closed Strings and Gravity}

As we mentioned before, superstring theory re-emerged in the 80s
after it had been realized that it provided a natural and consistent
host for gravitons\cite{Green:1984sg}. In order to be a bit more
specific, we shall consider closed strings propagating in some
background geometry $X$. It is widely known that superstrings
require $X$ to be 10-dimensional, so that contact with
4-dimensional physics is often made by rolling extra directions
up on small circles, or through more general compactifications.

Strings possess infinitely many vibrational modes which we can
think of as an infinite tower of massless and massive particles
propagating on $X$. The mass spectrum of the theory is linear,
with the separation that is parametrized by the tension $T_s \sim
1/\a'$ or, equivalently, by the length $l_s = \sqrt {\a'}$ of the
string. As strings propagate through $X$, they can interact by
joining and splitting. A simple such process for a one-loop
contribution to the $2 \to 2$ scattering of closed strings is
depicted in figure 1. Let us observe that any such diagram, no
matter how many external legs and loops it has, may be cut into
3-vertices. Consequently, all interactions between strings are
controlled by a single coupling constant $g_s$ that comes with the
3-vertex.

\begin{figure}[b]
\centerline{\psfig{file=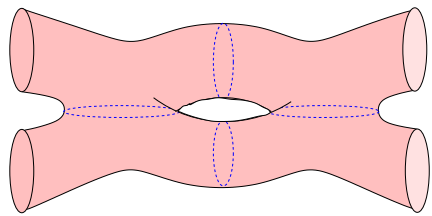,width=2.2in}}
\caption{Any string amplitude can be decomposed into 3-vertices.}
\label{fig1}
\end{figure}

String theory possesses a consistent set of rules and elaborate
computational tools to calculate scattering amplitudes. These
produce formulas of the form (\ref{Veneziano}). It is of
particular interest to study their low energy properties. When $E
\ll l^{-1}_s$, vibrational modes cannot be excited and all we see
are massless point-like objects. One may ask whether these behave
like any of the particles we know. The answer is widely known: at
low energies, massless closed string modes scatter like gravitons
and a bunch of other particles that form the particle content of
10-dimensional supergravity theories. This observation is
fundamental for string theory's advance into quantum gravity,
which came after failed attempts to develop perturbative quantum
gravity had led to the conclusion that Einstein's theory is
unlikely to be a fundamental theory of gravity. Similarly to
Fermi's theory of weak interactions, it should rather be
considered as an effective low energy theory that must be deformed
at high energies in order to be consistent with the principles of
quantum physics. String theory imposed itself as the most
promising candidate for a fundamental theory of gravity.

\subsection{Solitonic and D-Branes}

For a moment, let us turn our attention to (super-)gravity
theories. We are all familiar with the Schwarzschild solution of
Einstein's theory of gravity. It describes a black hole in our
4-dimensional world, i.e.\ a heavy object which is localized
somewhere in space. Similar solutions certainly exist for the
supergravity equations of motion. The massive (and charged)
objects they describe may but need not be point-like localized in
the 9-dimensional space. In fact, explicit solutions are known%
\cite{Duff:1994an} in which the mass density is localized
along p-dimensional surfaces with $p=1$ corresponding to strings,
$p=2$ to membranes etc. Such solutions were named black p-branes.
Like ordinary black holes, however, most of these objects decay.
But there exist certain extremal solutions, also known as
solitonic p-branes, that are stable.

Now let us recall from the previous subsection that supergravity
emerges as a low energy description of closed string theory.
Consequently, if supergravity contains massive $p+1$ dimensional
objects, the same should be true for closed string theory. One may
therefore begin to wonder about the role p-branes could play in
string theory. In order to gain some insight, let us suppose that
a brane has been placed into the 9-dimensional space of our string
background. Since it is heavy and charged, it will interact with
the closed string modes in this background. In supergravity, we
would describe this interaction through the exchange of gravitons
or other particles mediating the relevant interaction. In string
theory, a similar picture is possible only that now the
interaction is mediated by exchange of closed strings as shown on
the left hand side of figure 2.
\begin{figure}[b]
\centerline{\psfig{file=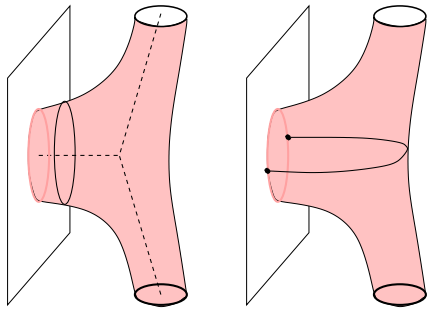,width=2.2in}}
\caption{There are two ways to think about the interaction between
closed strings and brane.}
\label{fig2}
\end{figure}
But the figure suggests another
way to think about the very same process. In order to allow for an
unbiased view, we have re-drawn the interaction process on the right
hand side of figure 2. What we see now is an infalling closed string
that seems to open up when it hits the brane. For a
brief period, an excited state is formed in which an open string
propagates with both its ends remaining attached to the brane.
Finally, this state decays again by emitting a closed string.
Hence, we found two very different ways to think about exactly the
same process. One of them involves an excited state of the p-brane
in which an open string travels along the p+1 dimensional
world-volume. In order for such a state to exist, branes in string
theory must be objects on which open strings can end. This is
indeed the defining  feature of so-called D(irichlet)p-branes in
string theory\cite{Polchinski:1996na}.

\subsection{D-Branes and Gauge Theory}

In the previous subsection we argued that D-brane excitations can
be thought of as open strings whose endpoints move within the
p-dimensional space of a brane. Therefore, branes provide us with
a second set of light objects, namely the vibrational modes of
open strings. One can ask again whether the massless open string
modes behave like any of the known particles. The answer is known
for a long time: When $E \ll l^{-1}_s$, massless open string modes
scatter like gauge bosons or certain types of matter.

\begin{figure}[b]
\centerline{\psfig{file=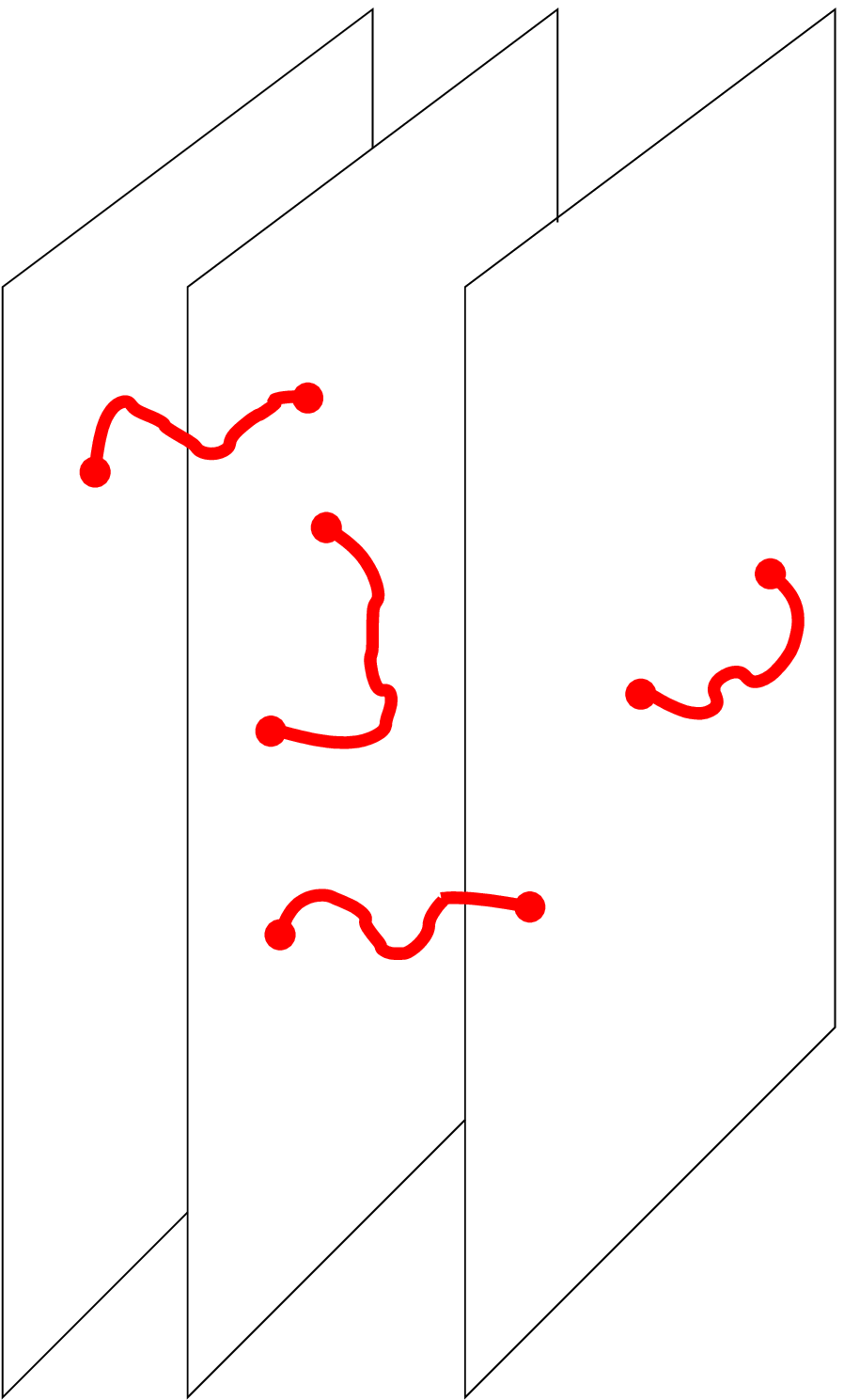,width=1.2in}}
\caption{Open strings can stretch between any pair of branes
in a stack. For bookkeeping purposes we introduce the color
indices $a,b$.}
\label{fig3}
\end{figure}

In order to obtain non-abelian gauge theories it is necessary to
consider clusters of branes. It is a remarkable fact of
supergravity that special clusters can give rise to stable
configurations. This is true in particular for a stack of $N$
parallel branes. Let us number the member branes of such a cluster
or stack by indices $a,b = 1, \dots, N$. Open strings must have
their end-points moving along one of these $N$ branes (see figure 3).
Since an open string has two ends, modes of an open string carry a pair
$a,b$ of `color' indices. Hence, massless open string modes on a
stack of $N$ parallel branes can be arranged in a $N \times N$
matrix, just as the components of a $U(N)$ Yang-Mills field. In
addition to non-abelian gauge bosons, various matter multiplets
can emerge from open strings. The precise matter content of the
resulting low energy theories depends much on the brane
configuration under consideration and we shall not make the
attempt to describe it in any more detail.

It is worth rehashing how $p+1$ dimensional gauge theories have entered
the stage through the back door. When we began this short cartoon
of string theory, closed strings (and therefore gravitons) were
all we had. Then be convinced ourselves that the theory contains
additional heavy $p+1$ dimensional D-branes. Their excitations
brought open strings into the picture and thereby another set of
light degrees of freedom, including non-abelian gauge bosons. Let
us stress once more that the latter do not propagate in the
10-dimensional space-time but rather on the $p+1$-dimensional
brane worlds. The dimension $p+1$ can take various values one of
them being $p+1=4$! Our sketch of modern string theory has now
brought us to the mid 1990s. At this point we have gathered all
the ingredients that are necessary to discover a novel set a of
equivalences between gauge and string theory.

\def\N{{\cal N}}
\section{AdS/CFT Correspondence}

We have reached the main part of this lecture in which we will
motivate and describe the celebrated AdS/CFT
correspondence\cite{Maldacena:1997re}.
Special attention will be paid to the simplest example of such a
duality between 4-dimensional $\N=4$ Super Yang-Mills (SYM) theory
and closed strings on $AdS_5 \times S^5$. This will enable us to
outline the formulation and the use of such dualities.

\subsection{String/Gauge Dualities}

The main origin of the novel dualities is not too difficult to
grasp if we cleverly combine what we have seen in the previous
part. To this end, let us suppose that we have placed two branes
in our 10-dimensional background and that they are separated by
some distance $\Delta y$. Since all branes are massive and charged
objects, they will interact with each other. In supergravity, we
would understand this interaction as an exchange of particles,
such as gravitons etc. Our branes, however, are objects in string
theory and hence there exists an infinite tower of vibrational
closed string modes modes that mediate the interaction between
them. A tree level exchange is shown on the left hand side of
figure 4. But as in our previous discussion, there exists another
way to think of exactly the same process in terms of open strings.
This is visualized on the right hand side of figure 4. There, the
interaction appears to originate from pair creation/annihilation
of open string modes with one end on each of the branes. In string
theory, these two prescriptions of the interaction give exactly
the same final result for the force between the two branes.

\begin{figure}[b]
\centerline{\psfig{file=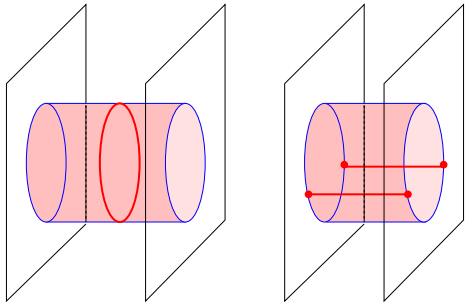,width=2.4in}}
\caption{There are two ways to think about the interaction between
two branes. One involves opens strings, the other is mediated by
closed string modes.}
\label{fig4}
\end{figure}

A closer look reveals that the equivalence of our two
computational schemes, one in terms of closed strings the other in
terms of open strings, is surprisingly non-trivial. Suppose, for
example, that the distance $\Delta y$ between the branes is very
large. Then the closed string modes have to propagate very far in
order to get from one brane to the other. Consequently,
contributions from massive string modes may be neglected and it is
sufficient to focus on massless closed string modes, i.e.\ on the
particles found in 10-dimensional supergravity. In the other
regime in which the separation between the two branes becomes of
the order of the string length $l_s$, such an approximation cannot
give the right answer. Instead, the full tower of closed string
modes must be taken into account. In other words, when $\Delta y
\sim l_s$ the supergravity approximation breaks down and we have
to carry out a full string theory computation. From the point of
view of open strings, the situation is reversed. When the branes
are far apart, pair created open strings only propagate briefly
before they annihilate again and hence the entire infinite tower
of open string modes contributes to this computation. In the
opposite regime where $\Delta y \sim l_s$, however, the
interaction may be approximated by restricting to massless open
string modes, i.e.\ all we need to perform is some gauge theory
computation.

As simple as these comments on figure 4 may seem, they lead to a
remarkable conclusion: In the regime $\Delta y \sim l_s$, some
calculation performed in the gauge theory on the world volume of
our branes should lead to the same result as a full fledged string
theory calculation for closed strings propagating in the
10-dimensional background. There are a few aspects of this
relation that deserve to be stressed. In fact, we observe that it
\begin{itemize}
\item
does neither preserve character nor difficulty of the computation,
\item
relates diagrams involving a different number of loops,
\item
relates two theories in different dimensions, i.e.\ it is
holographic.
\end{itemize}
These three features emerge clearly from our analysis. The first
point is obvious. In fact, the two computations are so different
that they would usually not be performed by members of the same
scientific community. Furthermore, in our example, we related a
gauge theory one-loop amplitude to a tree level diagram of closed
string theory, i.e.\ we showed that classical string theory
encodes information on quantum gauge theory and vice versa.
Finally, the gauge theory degrees of freedom are bound to the
$p+1$-dimensional world-volume of our branes whereas closed
strings can propagate freely in 9+1 dimensions. We shall see these
three features re-emerge in the concrete incarnations of the
gauge/string theory dualities we are about to discuss.

\subsection{$\N=4$ SYM theory \& $AdS_5$}

In the previous subsection we argued that string theory
should be able to produce fascinating novel relations between
gauge and string theory. But the picture was a bit too general to
fully appreciate the powerful implication of our discussion. In
order to be more specific, let us focus on the most studied
example.

It arises from a stack of $N$ parallel D3 branes
that are placed in a flat 10-dimensional (type IIB)
superstring background\cite{Maldacena:1997re}.
According to our general discussion, low energy excitations on
such a brane configuration are described be some 3+1-dimensional
gauge theory. The theory in question turns out to be an $\N=4$
Super Yang-Mills theory. In addition to the SU(N) gauge
bosons, this model possesses six scalar fields and a bunch of
fermions. Admittedly, except for being 4-dimensional, this is not
the most realistic model of our world. Not only does it possess
the wrong matter content, it also is an example of a conformal
field theory (CFT), i.e.\ it looks exactly the same on all length
scales, in sharp contrast to e.g.\ QCD. In particular, the $\N=4$ SYM
quantum theory has no confining phase. But for the moment we only
intend to explain some general ideas and so we defer such concerns
to the next section.

Gauge/string duality claims that, in the limit of large number $N$
of colors, $\N = 4$ SYM theory is dual to a theory of closed
strings which propagate in the curved near-horizon geometry of our
stack or D3 branes. The latter can be shown to split into the
product of a compact 5-sphere $S^5$ and a non-compact
5-dimensional Anti-deSitter (AdS) space $AdS_5$. Everybody knows
that $S^5$ consists of all points in 6-dimensional Euclidean space
that possess the same distance $R$ from the origin. $AdS_5$ can be
constructed in the same way only that the 6-dimensional Euclidean
space is now replaced by a space with 2 time-like and 4 space-like
coordinates,
$$
y^2_1 + y_2^2 + y_3^2 + y_4^2 - y_5^2 - y_6^2 = R^2 \ . \ \ \
\mbox{($AdS_5$)}
$$
The resulting 5-dimensional non-compact space includes one
coordinate $r$ that measures the radial distance from the stack of
branes in addition to the branes' four world-volume coordinates.

The assertion of the duality is that computations in $\N = 4$ SYM
theory and in string theory on $AdS_5 \times S^5$ give identical
results! Of course, gauge physicists and string theorists need to
use an extensive dictionary in order to compare the outcome of
their respective computations. We shall only discuss a few entries
of this dictionary here in order to give an idea of how this
works.

To begin with, let us talk about the parameters in the two
theories. On the gauge theory side, there are two of them, namely
the number $N$ of colors and the Yang-Mills coupling $g_{YM}$. In
the context of large $N$ limits, it is more appropriate to work
with the so-called 't Hooft coupling $\lambda = g_{YM}N^2$ instead
of $g_{YM}$. On the string theory side, we have the string coupling
$g_s$ and the radius $R/l_s$ of the $AdS_5$ space measured in
units of the string length $l_s$. We are prepared now to state the
first entry in the AdS/CFT dictionary which relates the two sets
of parameters as follows
\begin{equation} \label{parameters}
 \lambda \ = \ (R/l_s)^4 \ \ \ , \ \ \ \ 
N \ = \ \lambda \, g^{-1}_s\ \ \ . \end{equation}
Gauge theory computations are perturbative in $\lambda$ and hence
get mapped onto the extremely stringy regime in which the
curvature radius $R$ of $AdS_5$ is of the order of the string
length $l_s$. Furthermore, the comparison with perturbative string
theory results requires the string coupling $g_s$ to be small and
hence a large number of colors. We shall discuss this further in
the next subsection.

Let us turn to the second entry of the AdS/CFT dictionary. When
dealing with gauge theories, we are interested in gauge invariant
fields or operators, such as the stress-energy tensor, the trace
of the field strength etc. According to the AdS/CFT dictionary,
such operators correspond to the modes of closed strings moving in
$AdS_5 \times S^5$. One example of this map between gauge theory
operators and string modes involves the stress energy tensor of
the gauge theory which gets mapped to the massless graviton of the
closed string theory.

Listing gauge invariant operators in terms of closed string modes
may not seem such a big deal at first, until it is realized that
this map preserves additional data. We have mentioned above that
$\N=4$ SYM theory looks the same on all scales. Hence, re-scalings
can always be undone by a field re-definition. In this way, every
field is assigned its length dimension $\Delta$. In most cases,
the latter receives quantum corrections, i.e.\ it is given by the
classical dimension $\Delta_0$ of the field plus a quantum
contribution, the so-called anomalous dimension $\delta = \Delta -
\Delta_0$. With gauge invariant operators on the SYM side being in
one-to-one correspondence to closed string modes, one may now
wonder whether it is possible to determine anomalous dimensions
from string theory. AdS/CFT duality suggests that this is the case
and that the relevant quantity to compute is the mass of the
associated string mode. Let us test this quickly for the only
example we can treat without any effort: The stress energy tensor
of the gauge theory has a vanishing anomalous dimension. This
matches the fact that the graviton mode of the closed string is
massless. Even without carrying our discussion of the dictionary
through to any of its many further entries\cite{Maldacena:1997re},
we hope to have shown that the novel gauge/string dualities are
not only non-trivial but also quite concrete.

\subsection{Solution of the AdS/CFT}

In principle, the option to compute e.g.\ anomalous dimensions
from string theory opens very exciting new avenues. But there is
one issue with putting it to use right away: At this point, string
theory on $AdS_5$ has not been solved! In particular, we are not
able to write down its mass spectrum. In fact, while it is easy to
determine the spectrum of vibrational modes for strings in flat
space, solving the same problem for strings in the curved $AdS_5$
space is a very hard technical challenge. A helpful analogy is
provided by the spectrum of the Laplace operator. Finding its
eigenvalues on a torus required no work at all since eigenfunction
are simply plane waves. The same problem on a sphere or any other
curved space is significantly harder.

After this bad news there is a bit of good news too. String theory
on $AdS_5$ is solvable and there exists a certain amount of
technology already that deals with somewhat similar problems in
lower
dimensions\cite{Zamolodchikov:1995aa,Teschner:1997ft,Gotz:2006qp}.
Much more work will go into further developing the existing
methods of so-called integrable models, into the investigation of
their symmetries and various limiting regimes, before we can hope
to extract the desired information. But there is a growing number
of collaborations that have made this one of their prime tasks.

In the meantime, the AdS/CFT correspondence is far from being
useless. Let us note first that there exits a limit in which we
can estimate string theory quantities by their supergravity
approximations. As usual, this requires $l_s$ to be small or,
equivalently, $R/l_s$ to be large. Hence, according to eq.\
(\ref{parameters}), supergravity computations encode information
on strongly coupled gauge theory. This is certainly very exciting,
but it requires some faith into the correspondence, at least when
applied to gauge theory quantities that supersymmetry does not
protect from receiving quantum corrections. We shall come back to
some concrete supergravity predictions later on.

During the last years, a much more advanced approach was developed
that, for the first time, interpolates non-trivially between the regime
of perturbative gauge theory on one side and supergravity on the
other. It is based on the idea that, even before being able to
quantize string theory in $AdS_5$, we can determine its spectrum
in a semiclassical approximation when some quantum numbers become
large. To be a bit more specific, let us consider the anomalous
dimension $\delta = \delta(\lambda)$ of twist-2 operators in
the limit of large spin $S$ which can be shown to take the form,
\begin{equation}
\delta(\lambda) \ = \ f(\lambda) \, \log S + \dots
\end{equation}
where the $\dots$ stand for lower order terms in the spin $S$. The
universal scaling function $f(\lambda)$ multiplying the leading
$\log S$ term is also known as the
cusp anomalous dimension. Its behavior at large $\lambda$ was
first calculated from string theory a few years
ago\cite{Gubser:2002tv,Frolov:2002av}. In a beautiful development
initiated by Minahan and Zarembo\cite{Minahan:2002ve}, the function
$f(\lambda)$ has been determined recently by Beisert, Eden and
Staudacher\cite{Beisert:2006ez}. Their formula correctly
reproduces highly non-trivial gauge theory results up to four (!)
loops\cite{Moch:2004pa,Kotikov:2004er,Bern:2006ew} and then
interpolates all the way to large $\lambda$ where it matches
the strong coupling predictions\cite{Benna:2006nd}. To
date, this is certainly the most impressive demonstration of the
AdS/CFT correspondence. It required the use of highly non-trivial
technologies borrowed from integrable systems, in particular the
so-called Bethe-Ansatz that was introduced to solve problems in
statistical physics\cite{Bethe:1931hc}.

\section{Extensions and Applications}

In this final section we would like to briefly touch upon some of
the extensions and applications that go beyond the simple example
we have discussed so extensively above. This is an extremely active
field right now and therefore we are not be able to do it any
justice. In particular, we apologize to many authors of original
contributions for (almost) systematically avoiding references.

\subsection{More Realistic Models}

As we have pointed out before, the $\N =4$ SYM theory we have used
in the last section to illustrate the AdS/CFT correspondence is
far away from anything that resembles nature. In order for
gauge/string dualities to be of practical use, one needs to
construct examples in which the gauge theory has more realistic
features. These certainly include broken scale invariance, and in
particular confining models, (partially) broken super-symmetry, the
inclusion of finite temperature, flavor, chiral symmetry etc. The
search strategy is rather clear: Our
first example was obtained by placing D3-branes in a
10-dimensional flat background. In order to find new pairs of dual
theories we can modify both the background and the brane
configurations we start with. In this way, all the properties we
listed above have been realized in one way or another.

To address confinement, we should understand in some more detail
why string theory in $AdS_5 \times S^5$ cannot produce a confining
phase. Confinement is normally detected through a term in the quark
anti-quark potential that grows linearly with the separation. On the
dual string theory side we think of the quark anti-quark pair as being
located at $r=\infty$. The two gauge theory particles sit at the ends
of a string which
hangs deeply into the fifth dimension of $AdS_5$, pulled by the
gravitational attraction of the branes at $r=0$. For our
discussion of confinement it is a crucial observation that the
stack of D3 branes produces a gravitational red-shift factor
that vanishes at $r=0$, causing the string's effective tension
to approach zero in the vicinity of the branes. Hence, stretching
a string along the brane at $r=0$ costs no energy. This implies
that there is no linear term in the quark anti-quark potential
and hence there is no confinement. Consequently, we may think of
a confining background as one that is been capped off at some
finite $r = r_{min}$ close to the branes' location so the the
red-shift remains non-zero. Several concrete and basic
constructions of confining models\cite{Witten:1998zw,%
Polchinski:2000uf,Klebanov:2000hb} have been explored.

Studying gauge theories with less supersymmetry is possible if we
start our construction with a less supersymmetric background and/or
brane configuration. Similarly, we can heat the dual pair of theories
to finite temperature by placing a black hole into the 5-dimensional
geometry. Its temperature is felt in both string and gauge theory.
For lack of space we can neither go into any more detail nor
continue listing further QCD-like features and their string
theoretic implementation. The interested reader can find a
more satisfactory account of early constructions in a
nice review of Klebanov\cite{Klebanov:2005mh}. More recent
developments in this lively field run under
keywords such as holographic QCD or AdS/QCD.

\subsection{AdS/CFT and sQGP}

As described in the talk of X.N.\ Wang, there is evidence
that heavy ion collisions at RHIC produce a strongly coupled
quark gluon plasma (sQGP). From the point of view of string
theory such a discovery has some appeal. In fact, as we have
stressed several times before, the AdS/CFT correspondence
provides new leverage for advancement into strongly coupled
gauge physics and hence a sQCP could serve as an almost ideal
laboratory. Many characteristics of the plasma are discussed
in a rapidly growing number of publications on AdS/CFT and
sQCD, including shear viscosity, jet quenching etc. We are
not able to cover even a small fraction of these and limit
ourselves to comments on some early observations.

The QGP produced in heavy ion collision behaves like a liquid
that may be treated using concepts and methods of hydrodynamics.
In particular, by Kubo's formula, its viscosity $\eta$ is related
to correlations of the gauge theory's stress energy tensor through,
$$ \eta = \lim_{\omega \rightarrow 0} \frac{1}{2\omega}
\int dt d^3x  e^{i\omega t}  \langle [T(t,x), T(0,0)]\rangle .
$$
The right hand side of this equation should be evaluated at
the temperature of the plasma and for some finite gauge theory
coupling $\lambda$. Since the stress energy has spin $S=2$, a
string theory evaluation of Kubo's formula can only be performed
in the supergravity limit, i.e.\ at infinite 't Hooft coupling
$\lambda$. For the background of D3 branes, the result
is\cite{Policastro:2001yc}
\def\s{{\varsigma}}
\begin{equation}  \label{viscosity}
  (\eta /\s )^{\N = 4}_{\lambda = \infty} \ = \ 1/ 4\pi \ \ .
\end{equation}
Here, $\s$ denotes the entropy density of the plasma. Even
though the computation applies to the large color limit of $\N=4$
SYM at infinite coupling, rather than to usual QCD, the results
underestimates the observed values merely by a factor of two.
This seems particularly remarkable when compared to perturbative
gauge theory results which make predictions for small coupling
that are one order of magnitude higher than observations.
More realistic string theory models at finite coupling have
been argued to give numerical values for $\eta/\s$ that are
bounded from below by the result (\ref{viscosity}).

\subsection{High Energy scattering}

We do not want to conclude this introduction to the AdS/CFT
duality without explaining how the novel gauge string dualities
manage to circumvent what seemed like a `no-go' theorem in the
introduction. There we understood that the extended nature of
strings, as desirable as it is when modeling hadronic resonances,
causes cross sections to fall off way too fast at high energies.
The resolution is directly linked to the fact that strings in the
AdS/CFT correspondence possess a fifth dimension to propagate in,
namely the direction that parametrizes the distance $r$ from the
stack of branes. Consider an observer at infinity who is searching
for string modes at some given energy $E$, much smaller than the
string tension $T_s$. If the strings are far away from the brane,
the energy $E$ is not sufficient to excite the strings'
vibrational modes and all the observer sees are point-like objects
with the usual hard high energy scattering amplitudes. Strings
closer to the branes, however, appear red-shifted through the
branes' gravitational field. In other words, they possess an
effective tension $T_s^{eff}(r)$ that depends on $r$ and can
become so small that vibrational modes can be excited with the
energy $E$. The concrete from of the gravitational red-shift in
the D3 brane geometry implies that
\begin{equation}
 T_s^{eff}(r) \ = \ (r/R)^{2}\   T_s \ \ .
\end{equation}
Through this effect, our observer at $r=\infty$ is able
to see a tower of vibrational modes that are associated with
closed strings near $r=0$. Gauge theory amplitudes at small $s$
and $t > 0$ receive their dominant contribution from the region
near the branes' location and hence display the usual Regge
behavior with a Regge slope given by
$$ \a'_{eff} \ = \ \a' \ (R/r_{min})^2 \ \ . $$
The length $r_{min}$ was introduced in subsection 4.1 in the
context of our discussion of confining theories. For large center
of mass energy $s$ and $t < 0$, however, amplitudes are dominated
by string scattering processes near $r=\infty$ and hence they
possess the hard features of particle models\cite{Polchinski:2001tt},
in agreement with observation. Thereby we have explained how the
existence of a non-trivial fifth dimension for strings does indeed
overcome longstanding problems for the application of string theory
to strong interactions/gauge physics.

This brings us to the end of our short introduction to gauge/string
dualities. We have seen how they emerge from the modern picture of
string theory and were able to glimpse at some of their powerful and
concrete implications. Admittedly, much further work is necessary, in
particular to elevate our understanding of the relevant string theories
beyond supergravity and semiclassical approximations and, of course, to
get closer to studying real QCD. On the other hand, the path seems
very promising.

\end{document}